\documentclass[11pt]{article} 
\usepackage[utf8]{inputenc}
\usepackage{multirow}
\usepackage{amsfonts}
\usepackage{amsmath}
\usepackage{amssymb}
\usepackage{amsthm}
\usepackage{caption}
\usepackage[dvipsnames]{xcolor}
    \definecolor{darkgreen}{rgb}{0,0.5,0}
    \definecolor{darkblue}{rgb}{0,0,0.6}
    \definecolor{purple}{rgb}{0.4,.2,0.7}
\usepackage[margin = 2.5cm]{geometry}
\usepackage{graphicx}
\usepackage[hyperfootnotes = false, colorlinks = true, linkcolor = darkblue, citecolor = purple]{hyperref}
\usepackage{subcaption}

\newcommand{\be}{\begin{equation}}
\newcommand{\ee}{\end{equation}}
\newcommand{\bea}{\begin{eqnarray}}
\newcommand{\eea}{\end{eqnarray}}
\def\la{\label}
\def\nref#1{(\ref{#1})}
\def\half{{1 \over 2 }}

	\newcommand{\bes}{\begin{equation} \begin{split} }	
	\newcommand{\ees}{\end{split} \end{equation} }

	\usepackage{physics}

\begin{document}

\thispagestyle{empty}
\begin{center}
    ~\vspace{5mm}

  {\LARGE \bf {A simple quantum system that describes a black hole  \\}}

 %   {\bf   Juan Maldacena$^1$}

     %   $^1$Institute for Advanced Study,  Princeton, NJ 08540, USA 

   \vspace{0.5in}
     
   {\bf     Juan Maldacena %$^1$ 
   }

    \vspace{0.5in}
 
   ~
   \\
  % $^1$
  Institute for Advanced Study,  Princeton, NJ 08540, USA

    \vspace{0.5in}

    \vspace{0.5in}
    
%     {\tt   malda@ias.edu}

\end{center}

\vspace{0.5in}

\begin{abstract}

 During the past decades, theorists have been studying quantum mechanical systems that are believed to describe black holes. We review one of the simplest examples. It involves a collection of interacting oscillators and Majorana fermions. It is conjectured to describe a black hole in an emergent universe governed by Einstein equations. Based on previous numerical computations, we make an estimate of the necessary number of qubits necessary to see some black hole features.

 \end{abstract}

\vspace{1in}

\pagebreak

\setcounter{tocdepth}{3}
%{\hypersetup{linkcolor=black}\tableofcontents}

 \section{Introduction} 
 
The interesting advances in quantum simulators and quantum computers is opening up opportunities to study complex quantum systems whose behavior is hard to predict ahead of time. 
An interesting class of strongly interacting systems are those that could give rise to an emergent holographic universe: a spacetime governed by Einstein equations. 
 
We review here the ``simplest'' system that gives rise to a spacetime governed by Einstein gravity \cite{Banks:1996vh,Itzhaki:1998dd}.   It is simplest in the sense that it involves  ordinary quantum mechanics, instead of quantum field theory\footnote{There are simpler models, such as the SYK model \cite{Sachdev:1992fk,KitaevFirstTalk},  which reproduce many interesting features of gravity, but are not known to be described by a local bulk Lagrangian in the emergent spacetime, such as the one we have in the Einstein theory of gravity. These simpler models are thought to be related to a more exotic theory of gravity.}. By an Einstein gravity theory we mean a theory with a dynamical spacetime geometry governed by the Einstein Lagrangian $ S \propto \int \sqrt{g} R + \cdots $, where the dots indicate additional matter fields with a local spacetime Lagrangian. 

We first give a cartoon picture of the system in question, and then we give a more detailed description. Hopefully, no background in quantum gravity or string theory is necessary to understand this review.

\subsection{A very quick summary } 

%The simplest quantum system that we learn in quantum mechanics is two level system. Another simple system is a harmonic oscillator.  
%We can also consider a large collection of them. 

The quantum system  is constructed by starting with a collection of harmonic oscillators and Majorana fermions and then adding some special interactions among them. 
  The simplest   interaction would be a cubic potential. This has the disadvantage that it is unbounded below in some directions.  This is a problem if we want to make the coupling strong. For our purposes it is better to introduce a potential that is a perfect square, so that we can make sure that it is bounded below. 
In other words, something of the rough form 
\be \la{CarBos}
L_B \propto  \sum_A \left[   \dot x_A^2  +   \omega^2 x_A^2 + \left( \sum_{B,C} F_{ABC} x_B x_C \right)^2  ~\right] 
\ee 
where $F_{ABC} $ are some coupling constants, and the sum runs over all the oscillators. 
In addition, we also introduce interactions with the Majorana fermions 
\be 
L_F \propto  \sum_{A} \left[ \psi_A \dot \psi_A + 
% \sum_B i \omega \Omega_{AB} \psi_A \psi_B +
 i \sum_{BC} \tilde F_{ABC} x_A \psi_B \psi_C \right] 
\ee 
where 
%$\Omega_{AB}$ is an antisymmetric matrix with unit coefficients that gives rise to an energy splitting for the states of the fermions,  and 
$\tilde F_{ABC}$ is another set of coupling constants. 
The total lagrangian is the sum $L_B + L_F$. 

For the moment these are rather general interactions. The concrete system we will describe later has a very specific set of couplings, which are all fixed by  certain symmetries up to an overall coupling constant.  When this coupling is weak, we have a set of essentially independent oscillators. 
 On the other hand, when the coupling is taken to be large, the system is harder to describe. In fact, it is believed to give rise to an emergent spacetime.  For now we want to emphasize the simplicity of the setup: just some interacting bosonic and fermionic oscillators. 
 
\begin{figure}[t]
    \begin{center}
    \includegraphics[scale=.5]{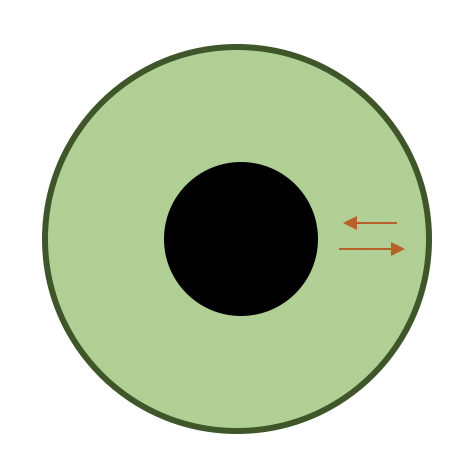}
    \end{center}
    \caption{A universe with a black hole in it. The gravitational potential grows without bound as we move away from the horizon. Therefore an excitation that comes out from the black hole (red arrow) would reflect back and fall in again.   }
    \label{BHB}
\end{figure}

 The emergent spacetime is simplest to describe at finite temperature, where it consists of a black hole in a``box universe".  By a ``box universe'' we mean a universe where the gravitational potential becomes large far from the black hole horizon, so that all outgoing excitations are reflected back to the black hole, see figure \ref{BHB}. 
 
 This universe is governed by Einstein's equations with suitable matter. The action includes gravity, a Maxwell or U(1) gauge field, and a few other fields. The black hole is electrically charged under the $U(1)$ gauge field. Both of these fields are emergent, they are unrelated to the gravitational or electric fields in our own four dimensional universe. Even though we start from a quantum system with no spatial extent, the emergent gravity system lives in a higher dimensional spacetime.    

Using these Einstein equations we can make a number of non-trivial computations which become predictions for the behavior of the quantum system at strong coupling, once we assume that the two related.
The simplest prediction is  for the relatively low energy thermodynamics. We find that  that the entropy scales with the temperature as $ S = \hat C T^{9/5}$ 
%\be 
% S \sim C T^{9/5 } ~,~~~~~~{\rm for } ~~~ F^{1/3}  \gg T \gg \omega 
 %\ee 
   with a specific constant $\hat C$. The factor of 9/5 can be viewed as a scaling exponent and is computed from the gravity solution. 
   
   In addition, there are numerous other observables whose behavior is predicted by the gravity description, such as correlation functions of certain operators. 
   
   It is interesting to compute the black hole quasinormal modes, which give us a ``spectrum'' of excitations around the black hole \cite{Biggs}. It is a spectrum in quotes because the quasinormal mode frequencies have an imaginary part representing the fact that the modes decay, or are falling into the black hole. These frequencies are signature of the black hole and they depend on the details of the geometry around the black hole. 
       
       For black holes in our universe, these quasinormal modes can be extracted from the LIGO/VIRGO gravity wave signals from black hole mergers \cite{Ghosh:2021mrv}, and the fact that they agree with the general relativity prediction is a very convincing piece of evidence that we are seeing black holes. 
  
   Similarly, seeing these quasinormal modes from a quantum simulation of the quantum system under discussion, would be a convincing evidence that we have created something that behaves as a black hole in the laboratory. 
   
  % There are also numerous other observables   whose behavior is predicted by the gravity description,  such as correlation functions of certain operators.  

   In the remainder of this article we give some more details on the statements made above. 
   
   \section{The quantum system } 
   \la{QS}
   
   In this section, we describe more precisely the quantum system that we are talking about. This quantum system is very special because it has many symmetries. 
   %Remember, it is just a collection of interacting oscillators. 
In fact, the structure of the interactions is completely determined by the symmetries. One of the symmetries   is $SU(N)$ and both the bosons and the fermions transform in the same representation, the adjoint. 

If we first set the frequencies of the oscillators to zero, we also have an $SO(9)$ symmetry.  
The bosonic degrees of freedom transform as a vector   of $SO(9)$. We can write them as 
 $X^{Ia}$ where $I=1, \cdots ,9$ is the $SO(9)$ vector index and    $a = 1, \cdots, N^2-1$ is the $SU(N)$ adjoint index. 
Then the interaction term takes the form 
\be \la{BosQu}
S_b  = \int dt \sum_{a =1}^{N^2-1} \left[   \sum_{I=1}^9 \half (\dot X^{a I } )^2 - {1 \over 4 }  {\lambda \over N} \sum_{I,J=1}^9   \left( \sum_{b,c=1}^{N^2-1} f^a_{\, \,bc} \, X^{Ia} X^{Jb}\right)^2  \right] 
  \ee 
  where $f^a_{\, \,bc}$ are the structure constants of $SU(N)$, which appear in the commutators of the generators of the algebra $[T_b,T_c]= i T_a f^a_{\,\, bc }$. 
   $\lambda$ is the coupling constant, and the factor of $N$ was introduced for convenience. Notice that $\lambda$ has dimensions of energy cubed as is the case for a cannonically normalized quartic oscillator such as in \nref{BosQu}. 
  An equivalent way to write \nref{BosQu} is the following. We can think of the $X^I$ as an $N\times N$ matrix (with zero trace), $(X^I)_i^{\,j}$, and write the kinetic and  interaction terms  as 
  \be \la{IntBos}
 S_B = {N \over \lambda } \int dt   \, {\rm Tr}\left\{ \sum_{I=1}^9 \half (\dot X^I)^2 + { 1 \over 4 }\sum_{I,J=1}^9       [ X^I, X^J ]^2 \right\} 
   \ee 
   where we rescaled the variables to pull out the coupling constant. 
 
  The fermions transform as a real 16 component spinor of $SO(9)$,  and are also in the adjoint of $SU(N)$, $\psi^{\alpha \, a}$, with $\alpha =1 , \cdots , 16$.  The interaction term then takes the form 
  \be \la{FerQu}
  S_F= { N \over \lambda } \int dt \left[ \half   \psi^{\alpha a } \dot \psi^{\alpha a }  +  { i \over 2  }     \psi^{\alpha \, a} \Gamma^I_{\alpha \beta}\psi^{\beta \, b} X^{I c} f^a_{\, \, bc} \right]
  = { N \over \lambda } \int dt Tr\left[ \half   \psi^{\alpha   } \dot \psi^{\alpha   }  +  \half           \psi^{\alpha } \Gamma^I_{\alpha \beta} [ \psi^\beta ,X^I ]  \right]
    \  \ee 
  where all repeated indices are summed.  In the second expression, we are thinking of $\psi^\alpha$ as matrix in the $SU(N)$ indices, $\psi^\alpha = ( \psi^\alpha)_j^{\, i } $. $\Gamma^I_{\alpha \beta}$ are real and symmetric $SO(9)$ gamma matrices, obeying $\{ \Gamma^I , \Gamma^J \} = 2 \delta^{IJ} $. 
 
  The relation between the overall coefficients in the interaction terms \nref{IntBos} and \nref{FerQu} is determined by another symmetry, called supersymmetry, which relates the bosonic and the fermionic degrees of freedom. We will not get into the details of the action of this symmetry. 
  The important point is that the full action $S_B + S_F$ is fixed by all these symmetries up to the overall coupling constant $\lambda$. This coupling constant has dimensions of energy cubed. Since the coupling is dimensionful, the only relevant quantity is the ratio between $\lambda$ and some other relevant energy scale. For example, when we consider the system at finite temperature the only dimensionless parameter is $T/\lambda^{1/3}$. 
 
This Lagrangian has a long and interesting history which we summarize in appendix A. Here we will be emphasizing only one aspect of the rich dynamics of this system.  

We can now reinstate the harmonic oscillator frequencies by adding the following terms to the action 
\be \la{MassTe}
S_\omega  = - { N \over \lambda } \int dt  Tr\left[ {  2 \omega^2 } \sum_{I=1}^3 (X^I)^2 +{ \omega^2 \over 2} \sum_{I=4}^9(X^I)^2  + { 3 i \over 4} \omega \psi  \Gamma^{1} \Gamma^{2} \Gamma^3 \psi  + 2 i  \omega \sum_{I,J,K=1}^3 \epsilon_{IJK} X^I X^I X^K \right] 
\ee 
which break $SO(9) \to SO(3) \times SO(6)$. $\epsilon_{IJK}$ is the ordinary epsilon tensor in three dimensions.  With these terms we see that the whole system, with action  $S_B + S_F + S_\omega$,  is describing just a collection of interacting  oscillators (both bosonic and fermionic). 
The terms in $S_\omega$ are also useful to eliminate some flat directions (or valleys) in the potential \nref{IntBos}, and have helped in some numerical computations \cite{Pateloudis:2022ijr}. For readers more at home in the Hamiltonian formalism, we give the full Hamiltonian in appendix B.

We will be interested in the parameter regime where 
\be \la{Range}
 \omega \ll T \ll \lambda^{1/3} ~,~~~~~~~~~N^{ - 5/9}\lambda^{1/3}  \ll T ~,~~~~~~~~N \gg 1
 \ee 
 %( ADD CONSTRAINT $ N^{-a} \ll \omega $, apparently it is $\lambda^{1/3} /N \ll T$.) 
 In this regime the system is strongly coupled. The reason is that the effective dimensionless coupling at temperature $T$ is given by 
 $\lambda/T^3$. This effective coupling is $N$ independent because  the factors of $N$ in \nref{IntBos} \nref{FerQu} were chosen so as to cancel factors of $N$ that arise because each of the degrees of freedom interacts with essentially $N$ other degrees of freedom. 
 At zero temperature, the effective coupling is $\lambda/\omega^3$.  
 For high temperatures, with $\lambda/T^3 \ll 1 $ the system is weakly coupled and not particularly interesting. The interesting behavior arises as the effective coupling becomes strong, $\lambda/T^3 \gg 1$.
 The second condition on the temperature in \nref{Range} is imposed so that the gravity solution we descibe below is valid. We can also consider lower temperatures \cite{Banks:1996vh} but we will not do it in this review. 
 
  % In this article, we will take the temperature to be large enough,   $N^{ - 5/9}\lambda^{1/3}  \ll T$
 
 In addition, we would like to restrict our attention to $SU(N)$ singlet states in the full Hilbert space. These singlet states  are supposed to  describe the gravity solution we discuss below. 
 % Regarding the $SU(N)$ symmetry is that we would like to restrict the states to the singlet sector. 
 This singlet restriction is equivalent to treating $SU(N)$ as a gauge symmetry. 
 It was argued in  \cite{Maldacena:2018vsr,Berkowitz:2018qhn}   that the non-singlet sector states have relatively higher energies at strong coupling, so that it should  not be necessary to explicitly impose the constraint; at low energies it seems to be imposed automatically.  It is also possible to effectively impose the constraint by adding terms to the Hamiltonian that are proportional to the square of the $SU(N)$ generators, so as to increase the energies of the non-singlet states.

 \section{The gravity solution} 
 
 The gravity solution is a certain black hole solution in a ten dimensional spacetime. When $\omega =0$,  it is a spherically symmetric solution.   The SO(9) symmetry is realized in gravity as the rotations on an 8 dimensional sphere.      
 It is a solution of an action that involves the Einstein term, a Maxwell term, and a scalar field 
 \be \la{GraAct}
 S = { 1 \over 16 \pi G_N} \int d^{10} x \sqrt{g} \left[ R - \half   (\nabla \phi )^2 - { e^{ { 3 \over 2 } \phi } \over 4 }F_{\mu \nu} F^{\mu \nu } + \cdots \right] 
 \ee 
 the dots indicate some other fields which are zero for the basic solution  we will discuss.  Notice that the scalar field sets the coupling of the Maxwell field.  
 
 This theory has various black hole solutions. The simplest one is the standard ten dimensional Schwarschild solution 
 \be 
 ds^2 = - f dt^2 + { dr^2 \over f } + r^2 d\Omega_8^2 ~,~~~~f = 1 - {r_s^7 \over r^7} ~,~~~~~~ F_{\mu \nu} =0 , ~~~\phi = {\rm constant} 
 \ee 
 where $d\Omega_8^2$ describes the unit radius metric on the eight dimensional sphere. This is very similar to the four dimensional Schwarzschild solution except that the two sphere is replaced by an eight sphere, and $1/r \to 1/r^7$ in the metric since this is the appropriate solution of the Laplace equation in nine spatial dimensions.  We would love to have a quantum mechanical description for this black hole, but we unfortunately do not. So, it is necessary to consider a slightly more complicated solution. 
 
 We  need to consider a charged black hole solution. Its charge is related to the electric flux on the sphere by the Gauss law
  \be \la{ElCh}
  N = { 1 \over 16 \pi G_N} \int_{S^8}  e^{{ 3 \over 2 } \phi} * F  =  { 1 \over 16 \pi G_N} \int_{S^8}  e^{{ 3 \over 2 } \phi} \vec E . d \vec S
   \ee 
where $N$ is an integer due to the charge quantization condition. This integer is 
  identified with the $N$ of the $SU(N)$ of the quantum system of section \ref{QS}. 
% The appearance of the scalar field in the action \nref{GraAct} implies that the scalar field will be non-constant.
 The general charged black hole solution of this kind is parametrized by a few parameters, the charge $N$, the mass, and the value of $\phi$ at infinity \cite{Horowitz:1991cd}. Due to the coupling between the electric flux and $\phi$ in \nref{GraAct} we find that $\phi$ is not constant.  
 
 We are interested in a particular limit of this black hole where the mass is close to its lowest possible value for a given charge $N$.
  We will not describe the precise way to take the limit.  We just give the final form of the   metric and scalar field (the gauge field is fixed by \nref{ElCh}).
 \bea \la{MetrF}
 ds^2 &=&  \sqrt{N}{ C^{1/8} \over 2 \pi }   \rho^{9/20} \left[ { 4 \over 25 } \left(- \rho^2  h d\tau^2 + { d\rho^2 \over \rho^2 h }\right)   + d\Omega_8^2 \right] ~,~~~~~~ e^{\phi } =  {  (2 \pi)^2 C^{3/4} \over N     \rho^{   21 \over 10} }
 \\ 
 &~& ~~h = 1 - \left({\rho_0  \over \rho }\right)^{14 \over 5 }  ~,~~~~~ ~ ~~C = 240 \pi^5 ~,~~~~\tau = { 5 \over 2} {  \lambda^{1/3}  t \over \sqrt{C} }  ~,~~~~~~~~G_{N} = 8 \pi^6 
 \eea 
 We have also set a convenient value for $G_N$. Here $t$ should be viewed as time normalized in the same way as in the quantum mechanical system. %There we have seen that $\lambda^{1/3}$ appears as a dimensionful coupling constant. 
 We see that everything depends only on $t \lambda^{1/3}$ as expected by dimensional analysis. Note that $\lambda$ here should be identified with the coupling constant in the quantum mechanical model.

We have also chosen a radial coordinate that makes it manifest that, for $\rho \gg \rho_0$,  the metric is proportional to $AdS_2 \times S^8$ up to an overall radial dependent function. $AdS_2$ is a two dimensional spacetime with constant negative curvature, whose metric can be written as $ds^2 = - \rho^2 d\tau^2 + d\rho^2/\rho^2 $.

 The particular exponents that appear in \nref{MetrF} depend on the details of the numerical coefficients in \nref{GraAct}. The other overall constants depend on the precise relation to the parameters of the matrix model. 
 
 There are a few noteworthy features of this metric. First notice that the overall factor of $g_{tt}$ starts being zero at the horizon, $\rho =\rho_0$,  and then rises monotonically when $\rho \to \infty$.  We can view this as a relativistic analog of a gravitational potential. The fact that it rises means that particles cannot escape to infinity. It is in this sense that the black hole is in a ``box", the walls of the (spherical) box are provided by this rising gravitational potential.  
 
 Note also that the metric in \nref{MetrF} has an overall factor of $\sqrt{N}$  which means that when we insert it into the action \nref{GraAct}, remembering we are in 10 dimensions, we get an overall factor of $N^2$. This means that $N^2$ is the relevant parameter that makes the gravitational theory weakly coupled. Namely, we need $N\gg 1 $ for the gravity theory to be weakly coupled.  Equivalently, we can say that the metric in \nref{MetrF} is essentially in Planck units. So we need a large $N$ for the space to be large in Planck units, or, more precisely, for the curvature to be small in Planck units.

 The parameter $\rho_0$ in \nref{MetrF} is given in terms of the temperature by the usual Hawking formula (adapted to this case)
 \be 
 { T \over \lambda^{1/3} } = { 7 \rho_0 \over 4 \pi \sqrt{C}  }   \la{Temp} \ee
As usual, the temperature is set by the condition that the euclidean version of the metric with periodic time  $t_E \sim t_E + {1 \over T} $ has no singularity at the horizon. 
 
 We can compute the entropy as a function of the temperature and we see that it is equal to 
 \be \la{EntP} 
 S = { {\rm Area } \over 4 G_N } = \tilde C    N^2 \left({ T \over \lambda^{1/3}} \right)^{9/5 } ~,~~~~~~~~~~~ \tilde C = 4^{13/5} 15^{2/5} \left( { \pi \over 7} \right)^{14/5}
 \ee 
From this, we can derive other thermodynamic quantities, such as the energy and the free energy. 
  We see that the entropy has a simple power law dependence on the temperature.
 In fact, the solution \nref{MetrF} displays an asymptotic scaling symmetry. Under $t\to \gamma t$ and $\rho \to \rho/\gamma$ the metric is rescaled by a factor of $\gamma^{-9/20}$. The full action gets an overall rescaling by $\gamma^{-9/5}$, which explains the temperature dependence in \nref{EntP}. Furthermore, since the action gets rescaled, this means that the equations of motion remain invariant. This means that physical observables have simple scaling properties. In other words, this is a critical system, developing a scaling behavior. It is not a true scaling symmetry of the quantum theory, but it is a scaling symmetry of the classical equations. From the point of view of the classical theory (large $N$ limit), it is as good as a scaling symmetry. In fact, correlation functions of certain local boundary perturbations display a power law behavior as a function of time, with $ \langle O(t) O(0) \rangle \propto t^{ - 2 \nu  } $ for some numbers $\nu $ which depend on the operator in question \cite{Sekino:1999av}. In other words, the gravity solution is telling us that the quantum system described in the previous section develops a peculiar critical behavior at strong coupling $\lambda/T^3 \gg 1$.   
 
 It is important to mention that the above action, as well as the metric we discussed, are a good description for $\rho$ small compared to one. This extra restriction, $\rho \ll 1$,  which is not apparent from \nref{MetrF}, is due to effects that are present in the full string theory description which we have not included here. Fortunately, when we are at relatively low temperatures $T \ll \lambda^{1/3}$, we see that $\rho_0 \ll 1$ and the horizon, together with  its environment, is in the region that we can indeed describe using just gravity.  This is the origin of the upper constraint on the temperature range in \nref{Range}.   This temperature regime where we can trust the metric is precisely the regime where the matrix model is strongly coupled.

 The solution we described so far is valid in the limit that $\omega \ll T$. The gravity solution for a more general situation, with  
   $\omega \sim T $,  is also known \cite{Costa:2014wya}, but we will not give the details here. It has a few more fields turned on.  In addition, there are phase transitions when $\omega \sim T$. For conceptual simplicity we have focused on the regime \nref{Range}, but for the purposes of quantum or classical simulation of black holes we can certainly contemplate also a regime with $\omega$ and $T$ which are comparable.  
 
 \subsection{Estimate for the number of qubits} 
 
 In \cite{Pateloudis:2022ijr}, the gravity prediction  \nref{EntP} for the entropy was compared against a numerical montecarlo\footnote{ The quantum mechanical model  has a sign problem due to the fermion determinant. In \cite{Berkowitz:2016jlq,Pateloudis:2022ijr} it was found that it was not severe in their range of temperatures.} computation in the quantum mechanical system for the following values of $T$ and $\omega$ and $N~$  \footnote{In  \cite{Pateloudis:2022ijr} they define $\mu_{\rm there} = 2 \omega_{\rm here}$.}
 \be \la{TempMu}
 { T \over \lambda^{1/3} } =0.3 ~,~~~~~~~~ {\omega \over T } = 0.8
~,~~~~~~~~~N = 16 
\ee 
 they found a result that only has a 13\% difference with \nref{EntP}, within the numerical error of the computation.      Notice that \nref{EntP} is a prediction with no free parameters. They also found that as $T/\lambda^{1/3}$ gets larger there are  larger deviations from gravity,  as expected.  For these values of $\omega/T$,  the corrections to the solution \nref{MetrF} are small \cite{Costa:2014wya}.   
 
 The parameters in \nref{TempMu} suggest the following very rough counting for the number of qubits for a quantum simulation of the model in a regime where we start getting agreement with gravity. 
 We have $8 N^2$ qubits from the $16 N^2$ Majorana fermions. For the bosons we have an infinite dimensional Hilbert space, but the important excitation levels are expected to be those up to $n \sim { \lambda^{1/3}/\omega }$. Therefore, we expect a number of qubits of the form 
 \be 
 n_q \sim N^2 \left[ 8 + 9 \log_2 \left({ \lambda^{1/3}\over \omega }\right)  \right] \sim 7,000 ~,~~~~~ {\rm for } ~~~  N=16 ~,~~~~~~~~ {\lambda^{1/3} \over \omega } \sim 4 
 \ee 
 This number is of the same order of magnitude as the number of logical qubits necessary for factoring integers faster than in a classical computer \cite{Preskill:2021apy}. Of course, one expects an error correction overhead. However, in our case, since we are interested in a finite temperature situation perhaps the error correction requirements are not so onerous. In other words, a quantum computer that can break RSA should probably also be able to simulate black holes!. For matrix models,  various simulation strategies, including quantum simulations,  were reviewed in \cite{Rinaldi:2021jbg}.
  
 Of course, a quantum simulation would test more observables, such as the prediction for quasinormal mode frequencies, correlation functions of operators, etc.   In addition, one could consider far from equilibrium questions such as the formation of the black hole, or its evaporation. More interestingly, it could offer some insights for how the geometry of the emergent spacetime is encoded in the quantum state of the quantum mechanical system.  
 
  We should emphasize that the ``universe'' that is described by the quantum system is effectively very small, its effective size in Planck units is not very large, certainly not as large as our universe!. One way to quantify the difference is to compare the entropy in our universe, which is of order of $S_{\rm our} \propto 10^{122}$, to the entropy  in \nref{EntP} which is about $S \sim 340 $ for the parameters in \nref{TempMu}\footnote{Another important difference is that our universe is expanding,  while the   universes that are generally believed to come from quantum systems are not expanding in the same way.}.
 
 Another issue that is relevant for quantum simulation is the number of gates or elementary operations that is needed\footnote{I thank A. Milekhin for raising this point.}. We can estimate this as follows. The Hamiltonian is the sum of many terms. The quartic term, \nref{IntBos},  contains $9^2 \times N^4 $ terms, with the $9^2$ coming from the sum over $I,J$ indices and the $N^4$ from the SU(N) indices. Similarly, the cubic term in \nref{FerQu} has $16^2 \times 9 \times N^3$ terms. These numbers add up to about $10^7$ for the parameters in \nref{TempMu}. The fermion operators lead to a further overhead of $\log_2 (16 N^2)$ \cite{Fermions},  if we need to express them in terms of qubits. For the bosonic creation and annihilation operators we also expect a significant overhead of order ${\lambda^{1/3}/\omega } $. These factors end up giving a number of order $10^8$. This is roughly the total number of gates necessary to implement the Hamiltonian.  To prepare the thermal state, we would need to apply the Hamiltonian for a number of time steps at least of order $\lambda^{1/3}/T$. The number of gates seems comparable to the number necessary to break RSA, at least with these very rough estimates.

 \section{Discussion} 
 
We have reviewed the connection between a relatively simple quantum mechanical system and a certain ten dimensional universe with a black hole in it. Let us now make a few remarks. 

\begin{itemize}
	\item 
We have not explained the reasons for the connection between the two.
 It was discovered through some studies of black holes in string theory. Going over the arguments for the connection would involve some details of string theory. The relation is still a conjecture, we do not have a mathematical proof for the equivalence between the two descriptions.

\item
The model might be amenable to analysis using other numerical methods, such as tensor network methods that have been useful for other theories with a sign problem. Bootstrap ideas were also recently explored \cite{Lin:2023owt}.  

\item
We can ask how much fine tuning is necessary to simulate this model. We can answer that by adding some other operators to the Lagrangian above. If the coefficients of the operators are sufficiently small, then we can analyze their fate using the gravity theory. As we said, the gravity theory develops a scaling behavior. This scaling can be used to classify operators in terms of their (anomalous) dimensions, which can be found using the gravity solution. It turns out that most operators get a relatively high anomalous dimension, so that even if they are present, they will not modify much the IR limit. Furthermore,  there are no ``relevant'' single trace SO(9) invariant operators. However,  there are some ``double trace'' SO(9) invariant operators, as well as several relevant non-SO(9) invariant operators. So, if we could somehow impose the SO(9) symmetry, or a large enough discrete subgroup, then it is possible that we could remove most of the relevant deformations, which is something that could help in getting to a simulation of the model discussed above. In particular, note that we want quartic interactions which are large, but we do not want six order interactions that are comparatively large.

% \item
% One feature of the dynamics that is believed to be important is the fact that the interactions are essentially all to all, in the sense that there is an  interaction term coupling any two of the oscillators. However, these interaction terms involve only at most four oscillators at a time. 
% There are other, simpler modes that can be considered with these features, such as the SYK model, which also have interesting strongly chaotic behavior and share many similarities with two dimensional gravity theories, or higher dimensional gravity theories around near extremal charged black holes.   
%In fact, SYK-like models are easier to simulate and there are already some initial results \cite{}. 
%What distinguishes the model described in this article from SYK is that here the bulk spacetime is described by a conventional gravity theory, with an Einstein Hilbert action, as opposed to a more complicated gravity theory. 

\item
This quantum mechanical model is developing an interesting many body state, a state that has an alternative interpretation in terms of a black hole in an emergent universe. We know that it arises in this model. However, we also like to  understand whether it arises also in other models, and quantum simulation might help identify other examples for which we do not have any conjectures. 

% \item
%In fact, we would love to understand in more detail how the geometrical description arises. We would like to understand how generic the phenomenon is, namely, does is happen with other models?. How much fine tuning is necessary? 

\item
There are other analogs of black holes, such as ``dumb holes'' \cite{Unruh:1980cg}, see \cite{Almeida:2022otk} for a review.  Those   systems capture interesting effects such as Hawking radiation, but not others such as the black hole entropy and the black hole microstates. Those analogs can be viewed as giving rise to quantum field theory in curved spacetime backgrounds  (possibly time dependent), but they do not appear to give rise to dynamical gravity governed by Einstein's equations.   What is special about the model described here is that it indeed gives rise to a dynamical spacetime governed by Einstein's equations. 

  \end{itemize}  
  
In conclusion, there is an interesting quantum mechanical model that has been conjectured to describe black holes. It seems to be difficult to simulate it with present quantum computers, but it should be possible with the ones that we are promised we will have in the not so distant future.

\subsection*{Acknowledgments}

We would like to thank the participants of the 2022 Solvay conference, including 
I. Cirac and M. Lukin, for discussions which prompted me to write this review. I also thank Anna Biggs, Jordan Cotler,  M. Hanada, A. Milekhin,  J. Santos and S. Shenker for discussions.   

J.M. is supported in part by U.S. Department of Energy grant DE-SC0009988.

\appendix

\section{The model in a broader context}

Here we give some historical comments and some pointers to the literature. 

The lagrangian we described above was first written in \cite{%Baake:1984ie,Flume:1984mn,
Claudson:1984th}, as an interesting quantum mechanical model due to its large number of supersymmetries. It was found as the dimensional reduction of ten dimensional super Yang Mills to just one dimension (the time direction). It was then used to analyze properties of quantized membranes in eleven dimensions  \cite{deWit:1988wri}. 

The most interesting application of this model  was found in  \cite{Banks:1996vh}. They proposed that a very low energy limit of the model can be used to compute scattering amplitudes in eleven dimensions. For this reason it is usually called the BFSS model. However, the energy regime necessary for the BFSS analysis is lower than the one discussed in this article. The BFSS energy regime  is actually {\it more} interesting than the one discussed this article,  since it would answer many questions about quantum gravity in eleven dimensions. However, it seems more difficult since it involves lower energies, energies parametrically small in the large $N$ limit.  The energy regime discussed in this article can be viewed as a stepping stone to the more challenging regime of BFSS. In addition, there are also interesting black hole questions already in this easier regime.  
    
     In \cite{Itzhaki:1998dd} the model was discussed in the energy regime discussed in this article, together with its gravity interpretation. The $\omega$ term \nref{MassTe} was added in \cite{Berenstein:2002jq}. 
  This model is also closely related to the so called AdS/CFT correspondence, or gauge/string duality  \cite{Maldacena:1997re,Gubser:1998bc,Witten:1998qj}. 
        
   The model has a very rich dynamics with various special  solvable  configurations. 
 There are numerous papers discussing features of this model. A {\it very} small sample of references is 
\cite{Boonstra:1998mp,Sekino:1999av,Polchinski:1999br,Taylor:2001vb,Kanitscheider:2008kd,Catterall:2009xn,Dong:2012se,Hanada:2013rga,Lin:2013jra,Asano:2014vba,Gur-Ari:2015rcq,Berkowitz:2016jlq,Pateloudis:2022ijr,Lin:2023owt}.

 \section{Hamiltonian } 
 
 The full Hamiltonian of the model is given by 
 \bea
 H &=& \sum_a \left[ \half \sum_{I=1}^9  p_{a I}^2   +{ \lambda \over N } {1\over 4 } \left( \sum_{b,c} \sum_{I,J=1}^9 f^a_{\, \, bc } x^{ I c } x^{ J b } \right)^2  + { \sqrt{ \lambda \over N } } \half i \sum_{\alpha \beta =1}^{16} \sum_{I=1}^9 \sum_{b,c}  \psi^{\alpha a} \Gamma^I_{\alpha \beta }  \psi^{\beta b} x^{ I c }f^a_{\, \, bc }  + \right. 
 \cr 
&~& + \left. \half \omega^2  \sum_{I=4}^9 x_{a I}^2 + \half (2 \omega)^2 \sum_{I=1}^3 x^2_{a I } + { 3 \over 4} i \omega \psi^{\alpha a} (\Gamma^1 \Gamma^2 \Gamma^3)_{\alpha \beta} \psi^{\beta a} - { \sqrt{ \lambda \over N } }  \,  \omega 
\sum_{I,J,K=1}^3 \sum_{bc} \epsilon_{IJK}  x^{a I} x^{b J} x^{c K} f^a_{\, \, bc} \right] ~~~~~~~~~~
\eea
where $\{ \psi^{\alpha a } , \psi^{\beta b} \} = 2\delta_{\alpha \beta} \delta_{ab}$ and $p_{aI}$ is the usual momentum conjugate to $x^{aI}$.  We took a basis of $SU(N)$ generators such that    $   Tr[T_a T_b] = \delta_{ab} $ and $[T_b,T_c] = i f^a_{\, \, bc} T_a $. The variables in this Hamiltonian have been normalized differently form the ones in the action in \nref{IntBos}, \nref{FerQu} and \nref{MassTe}.

\eject

\bibliographystyle{apsrev4-1long}
\bibliography{GeneralBibliography.bib}
\end{document}